\documentclass[twocolumn]{jpsj2}
\usepackage{graphicx}

\title{Anderson transition in the three dimensional symplectic
universality class}

\author{Yoichi Asada$^{1}$, Keith Slevin$^{1}$, and Tomi Ohtsuki$^{2}$}
\inst{$^{1}$Department of Physics, Graduate School of Science, 
Osaka University, 1-1 Machikaneyama, Toyonaka, Osaka 560-0043, Japan\\
$^{2}$Department of Physics, 
Sophia University, Kioi-cho 7-1, Chiyoda-ku, Tokyo 102-8554, Japan}

\date{\today}

\abst{
We study the Anderson transition in the SU(2) model
and the Ando model.
We report a new precise estimate of the critical exponent
for the symplectic universality class of the Anderson transition.
We also report numerical estimation of the $\beta$ function.
}

\kword{Anderson transition, symplectic class,
spin-orbit coupling, critical exponent, beta function}

\begin{document}

\maketitle

\section{Introduction}

The Anderson transition is a continuous
zero temperature quantum phase transition.
Just like continuous thermal phase transitions, such as
the magnetic phase transition in spin systems,
the critical phenomena of the Anderson transition are
 described by the single parameter scaling hypothesis.
\cite{abrahams:79}
It is thought that there are three universality
classes:
orthogonal, unitary, and symplectic.
In the absence of an applied magnetic field or other effect that
breaks time reversal symmetry,
disordered electron systems with spin-orbit coupling
belong to the symplectic universality class. 
Such systems are the subject of this proceeding.

One of the important quantitative characteristics of the Anderson transition 
is the critical exponent
$\nu$ that describes the divergence of the localization length $\xi$
\begin{equation}
 \xi \propto |x-x_c|^{-\nu}.
\end{equation}
Here $x$ is a parameter such as strength of disorder, Fermi energy,
and $x_c$ is its critical value.
For the orthogonal and unitary universality classes, $\nu$ 
has been determined
with high precision: $\nu=1.57\pm 0.02$ 
for the orthogonal class\cite{slevin:99}
and $\nu=1.43\pm 0.04$  for the unitary class \cite{slevin:97}
in three dimension (3D)
with $95\%$ confidence intervals.
By comparing the numerical estimates of the critical exponent in different models,
the universality of these values has also been tested  
\cite{slevin:97,slevin:99}.

The value of the critical exponent is expected to be helpful in
identifying the mechanism that drives the metal-insulator transition
observed in doped semiconductors and amorphous materials.
The theoretically estimated value $\nu$ can be compared
with the experimentally measured conductivity exponent $\mu$
\begin{equation}
 \sigma(T=0) \propto (x-x_c)^{\mu},
\end{equation}
through  Wegner's relation \cite{wegner:76}
$\mu=(d-2)\nu$, $d$ being the dimensionality of the system;
this predicts
$\mu=\nu$ in 3D.
Here $\sigma(T=0)$ is the conductivity at zero temperature.

In early work, the exponent $\mu$ was estimated
by extrapolating the conductivity to zero temperature.
More recent work avoids this unstable extrapolation by
using finite temperature scaling \cite{belitz:94}. 
This permits not only an estimation of the critical parameters
but also a check of the validity of the single parameter scaling theory.

Various estimates of the exponent $\mu$ have been obtained
experimentally (see Ref.~\citen{itoh:04} and references therein). 
Some
claim $\mu\approx 0.5$, which seems to
violate Chayes {\it et al.}'s inequality \cite{chayes:86,kramer:93} $\nu \geq 2/3$.
(Note that according to Wegner's relation $\mu=\nu$.)
Stupp {\it et al.} \cite{stupp:93,stupp:94} and Itoh {\it et al.} \cite{itoh:04}
pointed out that $\mu\approx 0.5$ is obtained
when data that may not be sufficiently close to the
critical point are analyzed.
When such data are excluded they found that $\mu \approx 1$
is always valid.

The value $\mu \approx 1$ is at variance with the theoretical values for
the 3D orthogonal and unitary classes.
To help clarify the origin of the discrepancy,
a precise theoretical estimate for the 3D symplectic class is desirable.
Furthermore, research in spintronics has sparked renewed interest in the effects of 
the spin-orbit interaction on quantum transport.\cite{zutic:04,murakami:03}

In this proceeding, we report precise analyses of
the critical phenomena at the Anderson transition
in the SU(2) model \cite{asada:02-1,asada:04-1}
and the three dimensional generalization of the
Ando model,\cite{ando:89,slevin:96}
both of which have spin-orbit coupling and have symplectic symmetry.
We also extend our finite size scaling analysis beyond the critical region
and present a numerical estimate of the $\beta$ function.

\section{Model and method}

The Hamiltonian used in this study describes
non--interacting electrons on a 
 square lattice with nearest neighbor hopping
\begin{eqnarray}
H=\sum_{i,\sigma}\epsilon_i c_{i\sigma}^{\dagger}c_{i\sigma}
-\sum_{\langle i,j \rangle,\sigma,\sigma'}
R(i,j)_{ \sigma \sigma'}
c_{i\sigma}^{\dagger}c_{j\sigma'}.
\label{eq:hamiltonian}
\end{eqnarray}
We distribute the random potential $\epsilon_i$
with box distribution
in the range $[-W/2,W/2]$.
Spin-orbit coupling is included in the $2\times 2$ matrix $R(i,j)$.

In the Ando model,
the hopping matrix depends only on the direction
\begin{eqnarray}
R(i,i+{\bf e}_n)&=&
e^{i\theta \sigma_n} \hspace{5mm} (n=x,y,z),
\end{eqnarray}
where ${\bf e}_n$ is the unit vector in the $n$ direction
and $\sigma_n$ is the Pauli spin matrix.
The parameter $\theta$, which is set to be $\theta=\pi/6$ in our
simulation, represents the strength of spin-orbit coupling.

In the SU(2) model,
the hopping matrix $R(i,j)$ is distributed randomly and independently
with uniform probability on the group SU(2)
according to the group invariant measure.
We parametrize the matrix $R(i,j)$ as
\begin{eqnarray}
  R(i,j)=
  \left(
  \begin{array}{cc}
  {\rm e}^{{\rm i}\alpha_{ij}} \cos \beta_{ij}
  & {\rm e}^{{\rm i}\gamma_{ij}} \sin \beta_{ij} \\
  -{\rm e}^{-{\rm i}\gamma_{ij}} \sin \beta_{ij}
  & {\rm e}^{-{\rm i}\alpha_{ij}} \cos \beta_{ij}
  \end{array}
  \right).
\end{eqnarray}
and distribute
$\alpha$ and $\gamma$ with uniform probability
in the range $[0,2\pi)$,
and $\beta$
according to the probability density,
$P(\beta){\rm d}\beta=
\sin (2\beta) {\rm d}\beta$  in the range $[0,\pi/2]$.

We consider a quasi-1D bar
of cross section $L^2$.
Periodic boundary conditions are imposed in the
transverse directions.
We calculate the quasi-1D localization length $\lambda$
with the transfer matrix method \cite{pichard:81,mackinnon:83}
and analyze the finite size scaling of
the dimensionless parameter $\Lambda=\lambda/L$.
In our simulation we fixed the Fermi energy $E=0$
and accumulated the data by varying disorder $W$ and system size $L$.

The single parameter scaling hypothesis implies
that $\Lambda$ obeys
\begin{equation}
 \ln\Lambda=F_{\pm}\left(\frac{L}{\xi}\right).
 \label{eq:spsout}
\end{equation}
Here $\xi$ is the correlation length in the metallic phase
and the localization length in the localized phase.
The subscript $\pm$ distinguishes the scaling function
in the metallic and localized phases.
When we analyze the critical phenomena of the Anderson transition,
it is more useful to use a different but equivalent form of the
single parameter scaling law
\begin{equation}
 \ln\Lambda=F_0(L^{1/\nu}\psi).
 \label{eq:spscr}
\end{equation}
Here $\psi$ is a smooth function of disorder $W$
that crosses zero linearly at the critical point $W=W_c$.

Single parameter scaling of $\Lambda$ can be described by
the $\beta$ function\cite{mackinnon:83,asada:04-1}
\begin{equation}
 \beta(\ln\Lambda)=\frac{{\mathrm d}\ln \Lambda}{{\mathrm d}\ln L}
 \label{eq:betafunc}.
\end{equation}
We estimate the $\beta$ function from the finite size scaling
analyses of numerical data.
The $\beta$ function in the critical region is calculated
from $F_0$ and the critical exponent $\nu$, and outside
the critical region from $F_{\pm}$.

\section{Critical phenomena of the Anderson transition}
\begin{table*}[tb]
\caption{\label{table:results1}
The details of various analyses where
irrelevant corrections to single parameter scaling are considered,
and the best fit estimates of the critical parameters.
$N_d$ is the number of data used in each analysis,
$N_p$ is the number of parameters,
$Q$ is the goodness of fit probability.
The precision is expressed as $95\%$ confidence intervals.
}
\begin{tabular}{lllllllllllll}
\hline
Model & $L$ & $N_d$ & $n_R$ & $n_I$ & $m_R$ & $m_I$ & $N_p$ &  $Q$
& $W_c$ & $ \ln \Lambda_c$ & $\nu$ & $y$
\\ \hline
SU(2) & [4,14] & 279 & 3 & 1 & 2 & 0 & 10 & 0.92 & $20.001 \pm .017$ & $-0.616\pm .005$
& $1.375 \pm .016$ & $-2.5 \pm .8$  \\
Ando  & [4,14] & 279 & 3 & 1 & 2 & 0 & 10 & 0.9  & $19.099 \pm .009$&$-0.605\pm .002$ 
& $1.360 \pm .006$ & $-3.8\pm .4$ \\
\hline
\end{tabular}
\end{table*}

\begin{table*}[tb]
\caption{\label{table:results2}
The details of analyses where corrections
to scaling are neglected.
The precision is again expressed as $95\%$ confidence intervals.
}
\begin{tabular}{llllllllll}
\hline
Model & $L$ & $N_d$ & $n_R$ & $m_R$ & $N_p$ &  $Q$
& $W_c$ & $ \ln \Lambda_c$ & $\nu$ 
\\ \hline
SU(2) & [8,14] & 156 & 3 & 2 & 7 & 0.7  & $19.984 \pm .008$ &$-0.612\pm .002$ 
& $1.367 \pm .007$ \\
Ando & [10,14] & 115 & 3 & 2 & 7 & 0.6  & $19.092 \pm .013$ &$-0.603\pm .003$ 
& $1.361 \pm .011$ \\
\hline
\end{tabular}
\end{table*}

\begin{figure}[tb]
\includegraphics[width=\linewidth]{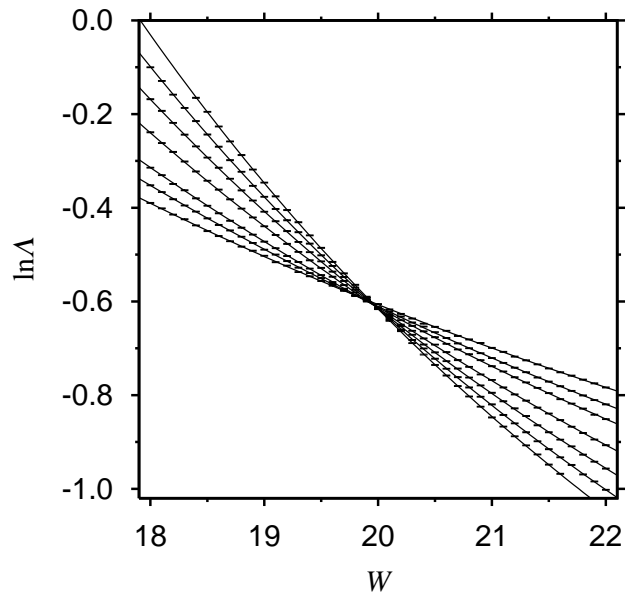}
\caption{
$\ln \Lambda$ as a function of disorder $W$
for the SU(2) model.
The lines are the fit to the function
$(n_R,n_I,m_R,m_I)=(3,1,2,0)$.
Although it may be difficult to confirm the existence of
corrections to single parameter scaling in this figure,
finite size scaling fit shows that such corrections do exist.
}
\label{fig:su2cp}
\end{figure}

\begin{figure}[tb]
\includegraphics[width=\linewidth]{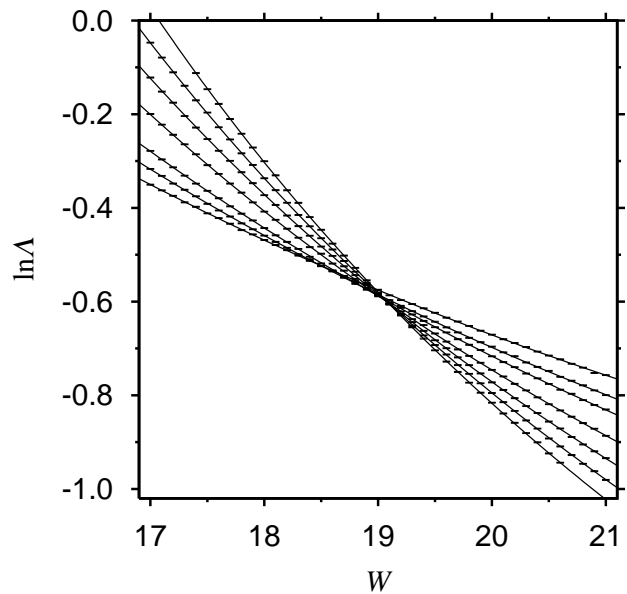}
\caption{
$\ln \Lambda$ as a function of disorder $W$
for the Ando model.
The lines are the fit to the function
$(n_R,n_I,m_R,m_I)=(3,1,2,0)$.
There are relatively larger irrelevant corrections
to single parameter scaling.
}
\label{fig:andocp}
\end{figure}

We calculate $\Lambda$
for sizes $L=4,5,6,8,10,12,14$ with
an accuracy of $0.1\%$.
The numerical data and the associated fits are shown
in Figs. \ref{fig:su2cp} and \ref{fig:andocp}.
In the absence of any irrelevant corrections to single
parameter scaling Eq.~(\ref{eq:spscr}),
plotting $\Lambda$ vs $W$ should show the critical
disorder as the common crossing point of the data.
However, the curves for different sizes
in Figs.~\ref{fig:su2cp} and \ref{fig:andocp}
do not cross at a common point, indicating the existence
of irrelevant corrections to single parameter scaling.
\cite{slevin:99}
Such corrections to perfect single parameter scaling
are usually present in a simulation of a finite system.

To analyse data over the full range of system sizes $L=[4,14]$,
we must take account of corrections to single parameter scaling
due to an irrelevant variable
\cite{slevin:99},
\begin{equation}
 \ln \Lambda= F_0\left(\psi L^{1/\nu}\right)
+\phi L^yF_1\left(\psi L^{1/\nu}\right).
\end{equation}
where $y$ is an irrelevant exponent
and $\phi L^y$ is the
corresponding irrelevant scaling variable.
For the purpose of fitting,
we approximate $F_0$ and $F_1$ by their finite order expansions
\begin{eqnarray}
 F_0(x)&=& \ln\Lambda_c + x + a_2 x^2 + \cdots + a_{n_R}x^{n_R} \\
 F_1(x)&=& 1+b_1 x +\cdots + b_{n_I}x^{n_I}.
\end{eqnarray}
We also expand $\psi$ and $\phi$
in terms of the dimensionless disorder $w=(W_c-W)/W_c$,
\begin{eqnarray}
 \psi&=&
\psi_1 w +\psi_2 w^2 +\cdots + \psi_{m_R} w^{m_R} \\
 \phi&=&
\phi_0 + \phi_1 w +\cdots + \phi_{m_I} w^{m_I}.
\end{eqnarray}
The details and results of the analyses
are presented in Table~\ref{table:results1}.

By discarding data for smaller systems, an analysis of the data
that neglects irrelevant corrections becomes possible.
A reasonable fit is obtained in the SU(2) model for $L\geq 8$
and in the Ando model for $L\geq 10$ (Table~\ref{table:results2}).
The estimated critical parameters in Table~\ref{table:results2}
are in reasonable agreement
with those in Table~\ref{table:results1}.

The most important point to be drawn from Table~\ref{table:results1}
and Table~\ref{table:results2} is that the estimates of the exponent
$\nu$ for the SU(2) model and the Ando model are in almost perfect
agreement.
As for the irrelevant exponent $y$,
our estimates are much less precise and
we can not clarify whether the values of $y$ for these two models
are the same or not.

Figures \ref{fig:su2cp} and \ref{fig:andocp} show that
the movement of the crossing points in the SU(2) model is smaller
than that in the Ando model,
indicating that the magnitude of irrelevant corrections
in the SU(2) model is smaller.
At the same time, we can expect that
the spin relaxation length
is shorter in the SU(2) model because of the random spin-orbit
coupling.
This suggests that
the leading irrelevant corrections observed for small $L$
in the Ando model might be from its relatively
larger spin relaxation length.

\section{Estimation of the $\beta$ function}

\begin{figure}[tb]
\includegraphics[width=\linewidth]{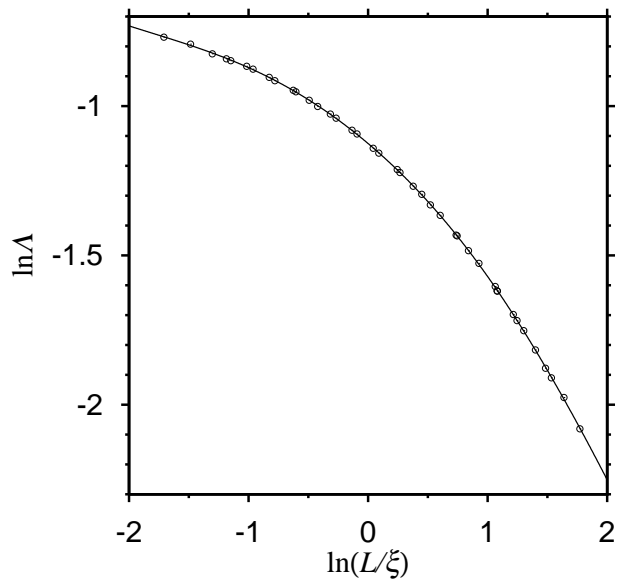}
\caption{
Single parameter scaling in the localized region
is demonstrated.
}
\label{fig:su2spsinsulator}
\end{figure}

\begin{figure}[tb]
\includegraphics[width=\linewidth]{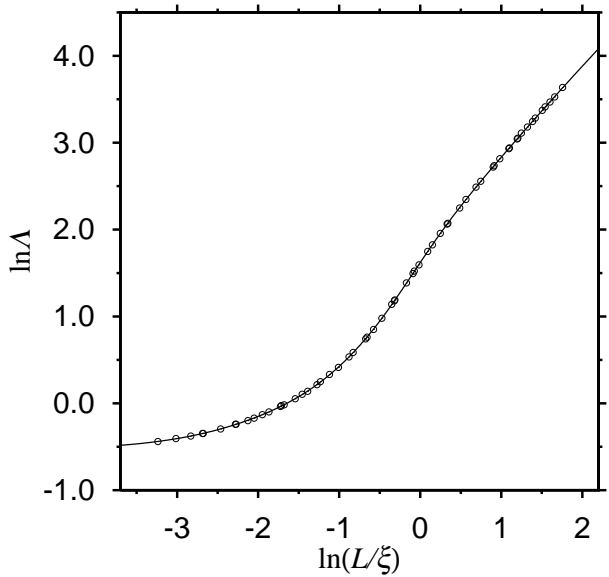}
\caption{
Single parameter scaling in the metallic region
is demonstrated.
}
\label{fig:su2spsmetal}
\end{figure}

\begin{figure}[tb]
\includegraphics[width=\linewidth]{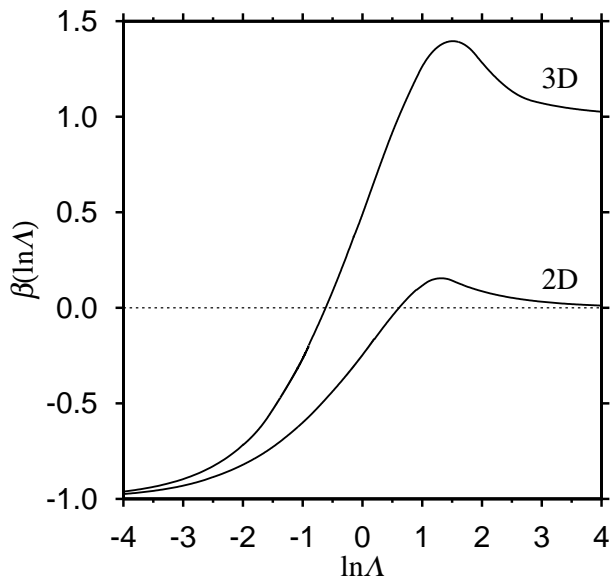}
\caption{
The $\beta$ functions for $\Lambda$.
The $\beta$ function in 2D is from Ref.~\citen{asada:04-1}.
}
\label{fig:betafunc}
\end{figure}

We estimate the $\beta$ function Eq.~(\ref{eq:betafunc})
by accumulating more
numerical data in the metallic and localized regions
of the SU(2) model at $E=0$.
The ranges of disorder and system size are
$W=[4,19]$ and $L=[8,14]$ for the metallic phase,
$W=[21,30]$ and $L=[8,16]$ for the localized phase.
The precision of the data for $\Lambda$ is $0.3\%$ or $0.6\%$.
The data covers the strongly localized 
to the strongly metallic limits.

We perform five scaling analyses for five regions:
strongly localized, localized, critical,
metallic, and strongly metallic.
From these analyses we estimate the $\beta$ function.
The methods of analyses are almost the same as those
used in 2D \cite{asada:04-1},
so we do not describe them in detail here.
We mention only one difference in the strongly metallic region.
As discussed in Ref.~\citen{mackinnon:83},
the asymptotic value of $\beta(\ln\Lambda)$
may depend on the dimensionality $d$
\begin{equation}
 \beta(\ln\Lambda) \rightarrow d-2 \hspace{5mm}(\ln\Lambda\rightarrow\infty).
\end{equation}
We analyze the data in the strongly metallic region
supposing that the asymptotic value is one in 3D.
For large but finite $\Lambda$ we speculate
that derivations from the asymptotic value can be described
by an expansion in powers of $\Lambda^{-1}$.
Stopping at the first order we have
\begin{equation}
 \beta(\ln \Lambda)=1+\frac{b}{\Lambda} \hspace{5mm}(\Lambda \gg 1).
\end{equation}
This corresponds to a linear increase of $\Lambda$ with $L$
\begin{equation}
 \Lambda=a\frac{L}{\xi}-b \hspace{5mm} (\Lambda\gg 1).
\end{equation}
Numerical data for large $\Lambda$ are well fitted by this form.

In Figs. \ref{fig:su2spsinsulator} and \ref{fig:su2spsmetal}
we demonstrate the single parameter scaling
in the localized and metallic region.
We can see from these figures that
the data for different values of disorder $W$ and
system size $L$ fall on a common scaling curve
when expressed as a function of $L/\xi$.

The resulting $\beta$ function
is displayed in Fig.~\ref{fig:betafunc}
as well as the $\beta$ function in 2D,
which is reported in Ref.~\citen{asada:04-1}.
The $\beta$ functions show non-monotonic behavior.
The maximum values are
$\beta_{\mathrm max}\approx 1.40$ in 3D and
$\beta_{\mathrm max}\approx 0.15$ in 2D.

\section{Summary and Discussion}

In summary,
we analyzed the scaling of the parameter $\Lambda$ in 3D.
The critical phenomena of the Anderson transition
in the SU(2) model and the Ando model are analyzed
and the critical exponent is estimated with high precision.
All the estimates of the critical exponent are
in the range $\nu=[1.35,1.39]$.
Our estimate is consistent with the previous estimates
$\nu=1.3\pm 0.2$ \cite{kawarabayashi:96} and
$\nu=1.36\pm 0.10$ \cite{hofstetter:98}
for the symplectic class.
We also estimated the $\beta$ function over the full range
from the localized to the metallic limits.

Our precise estimate clearly distinguishes the value of the
critical exponent $\nu$ in the symplectic class
from that for the orthogonal class
$\nu=1.58 \pm 0.02$. \cite{slevin:99}
On the other hand, comparing with the available estimate\cite{slevin:97}
$\nu=1.43\pm 0.04$  for the unitary class,
the precision is
not yet  sufficient to allow the same statement to be made, 
although the theoretical expectation is that the values for the 
unitary and symplectic classes should be different,
as was clearly demonstrated in 2D.\cite{asada:02-1,asada:04-1,huckestein:95}

The values of the critical exponent for three universality
classes of the Anderson transition
are inconsistent with the exponent $\mu\approx 1$ measured
in experiment.
This may indicate that the observed transition is not a 
pure Anderson transition and that
electron-electron interaction effects may have to be taken into account to
properly account for the critical phenomena
\cite{belitz:94}.


\end{document}